\documentclass[review]{elsarticle}
\usepackage{color,soul}
\usepackage{lineno,hyperref}
\modulolinenumbers[5]

\journal{Journal of Magnetism and Magnetic Materials}

\begin{document}

\begin{frontmatter}

\title{A strategic high throughput search for identifying stable Li based half Heusler alloys for spintronics applications}

\author{Rohit Pathak\fnref{myfootnote}}
\author{Parul R. Raghuvanshi}
\author{Amrita Bhattacharya\corref{mycorrespondingauthor}}
\address{Ab initio Materials Simulation Laboratory, Department of metallurgical engineering and materials science, Indian Institute of Technology, Bombay, Powai-400076, Mumbai, Maharashtra, India.}
\fntext[myfootnote]{rohitphysics137@gmail.com}
\cortext[mycorrespondingauthor]{Corresponding author}
\ead{b\_amrita@iitb.ac.in}

\begin{abstract}
In this work, high throughput DFT calculations are performed on the alkali metal-based half Heusler alloys; LiY$_p$Y$^\prime_{1-p}$S (Y, Y$^\prime$ = V, Cr, Mn, Fe, Co, Ni and \textit{p} = 0, 0.25, 0.5, 0.75, 1). Starting with 243 structural replica, systematic filters are designed to select the energetically and vibrationally favorable compositions by considering the contributions stemming from the magnetic alignments of the ions. Thereby, 26 dynamically stable magnetic compositions are identified, of which 10 are found to be ferromagnetic (FM), 4 antiferromagnetic (AFM) and 12 ferrimagnetic (FiM). 4 FM and 8 FiM ones are found to show 100 \% spin polarization. Further, tetragonal distortion is found to be present in 4 FM, 3 FiM and 4 AFM compositions, which indicates the possibility of easy-axis magnetocrystalline anisotropy. The ferromagnetic LiFe$_{0.5}$Mn$_{0.5}$S and antiferromagnetic LiFeS are found to have the most prominent easy-axis magnetocrystalline anisotropy.
\end{abstract}


\end{frontmatter}


\section{Introduction}
Heusler alloys comprise an exciting class of materials exhibiting versatile properties \cite{felser2015heusler, webster1969heusler, poon2001electronic, lin2010half, gruhn2010comparative, bai2012data, xiao2010half, chadov2010tunable, ma2017computational, graf2010heusler, nanda2003electronic, felser2015basics, casper2012half, helmholdt1984magnetic} ranging from half metallicity\cite{wurmehl2006half, felser2015basics}, superconductivity\cite{graf2010heusler, graf2011simple}, magnetic properties\cite{wang2014heusler, webster1969heusler} etc. Naturally, they have been used extensively in tunnel junction devices\cite{faleev2017heusler, felser2013tetragonal, felser2013spintronics}, solar cells\cite{gruhn2010comparative, graf2010heusler}, thermoelectrics\cite{poon2001electronic, raghuvanshi2020high}, and spintronics based applications\cite{bader2010spintronics, vzutic2004spintronics}. Since its discovery in the year 1903, this alloy class has an ever-escalating demand in the material science research community \cite{graf2011simple, elphick2021heusler, ma2017computational, faleev2017origin, fang2002spin, kandpal2007calculated, oliynyk2016high, faleev2017heusler}. Heusler alloys can be further majorly categorized as full Heusler (X$_{2}$YZ)  and half Heusler (XYZ) depending upon the full or half occupancy of X site elements in the lattice \cite{graf2011simple, helmholdt1984magnetic, felser2015heusler, graf2010heusler}. For spintronics applications, Heusler alloys, with X and Y elements from transition metal family and Z element from the main group, have been reported to be particularly promising \cite{bader2010spintronics,li2016first, chambers2003new, vzutic2004spintronics, palmstrom2016heusler, ma2017computational, ma2018computational, felser2013tetragonal, guillemard2019ultralow, meinert2010ab, balke2008rational}. These materials are found to exhibit the key properties necessary for spintronics applications viz. half metallicity, high magnetization, and uniaxial magnetocrystalline
anisotropy \cite{chambers2003new, elphick2021heusler, vzutic2004spintronics, bader2010spintronics}. Apart from theoretical predictions \cite{li2016first, gao2019high, ma2017computational, ma2018computational}, there are plenty of reports of their experimental synthesis \cite{graf2010heusler, oliynyk2016high, helmholdt1984magnetic, wang2014heusler, ahmad2019size, ahmad2021first, kojima2019emergence, galdun2018intermetallic, balke2007mn}. However, it is impossible to experimentally tune the elemental constituents stoichiometrically
to achieve the best one with improved properties pertaining to the enormous compositional space. In this regards, density functional theory (DFT) has emerged as a powerful computational tool for materials discovery, which, if implemented in high throughput loops, not only leads to the identification of the efficient composition but also sheds light on the underlying phenomena \cite{oliynyk2016high, ma2017computational, kandpal2007calculated, faleev2017heusler, gao2019high, balluff2017high, li2016first}. As for instance, Faleev \textit{et al.} performed high throughput \textit{ab initio} DFT based study on 3\textit{d} and 4\textit{d} transition metal (TM) based full Heusler alloys (X$_2$YZ) with the Z element from the main group (i.e. from group III, IV, and V)\cite{faleev2017heusler}. Their search lead to the prediction of $\sim$ 19 potential candidates with high spin polarization and magnetocrystalline anisotropy. Ma \textit{et al.} performed a high throughput \textit{ab initio} DFT based study on transition metal (TM) based half Heusler alloys, XYZ (with X = Cr, Mn, Fe, Co, Ni, Ru, Rh; Y = Ti, V, Cr, Mn, Fe, Ni; Z = Al, Ga, In, Si, Ge, Sn, P, As, Sb) \cite{ma2017computational}. In their high throughput search they predicted several semiconducting ($\sim$ 26), half-metallic ($\sim$ 45) and near half-metallic ($\sim$ 34) half Heusler alloys with negative formation energy. Recently Amudhavalli \textit{et al.} explored Mn-based half Heusler alloys via DFT based first principle simulations, XYZ (X = Ir, Pt, Au; Y = Mn; Z = Sn, Sb) and predicted the half-metallicity of all alloys at high pressure\cite{amudhavalli2017structural}. 

While the TM based Heusler alloys are typically the most explored ones \cite{li2016first, gao2019high, ma2017computational, ma2018computational,graf2010heusler, oliynyk2016high, helmholdt1984magnetic, wang2014heusler, ahmad2019size, ahmad2021first, kojima2019emergence, galdun2018intermetallic, balke2007mn}, recent studies reveal spintronics potential of Alkali metal based Heusler alloys. Zhang \textit{et al.} studied Li-based half Heusler alloy viz. LiCrS and predicted one of its metastable phase possess a significant magnetic moment with the Cr atom showing atomic-like moment of 5 $\mu_B$ \cite{zhang2016half}. Shakil \textit{et al.} studied the magnetic and elastic properties of LiCrZ (Z = P, As, Bi and Sb) half Heusler alloys \cite{shakil2019theoretical} and found these alloys to be 100\% spin-polarized with very high magnetization. Recent study of Wang \textit{et al.} on alkali metal-based XCrZ (with X = Li, K, Rb, and Z = S, Se, Te) half Heusler alloys \cite{wang2017largest} revealed robust half metallicity (HM) in almost all alloys in their stable phases except for LiCrZ, which shows HM nature in only metastable phase. Alkali metal-based Heusler alloys have also been explored in the context of thermoelectric materials; for instance, Yadav \textit{et al.} investigated LiYZ (Y = Be, Mg, Zn, Cd and Z = N, P, As, Sb, Bi) series and found that LiZnSb has a typically large n-type power factor \cite{yadav2015first}. Although these studies hint towards the promise of Alkali metal based Heusler alloys for spintronics applications, but a detailed systematic analysis of magnetic properties of Alkali and transition metal based half Heusler alloys are hitherto absent in the literature.  

In this work, high throughput DFT calculations are performed on LiY$_p$Y$^\prime_{1-p}$S (Y, Y$^\prime$ = V, Cr, Mn, Fe, Co, Ni and \textit{p} = 0, 0.25, 0.5, 0.75, 1) compositions to identify the electronically and vibrationally stable ones for their spintronics applications.

\section{Methodology}
\label{metho}
The conventional cell of a full Heusler alloy (X$_2$YZ) has a cubic lattice with  4a (0, 0, 0), 4b (0.5, 0.5, 0.5), 4c (0.25, 0.25, 0.25) and 4d (0.75, 0.75, 0.75) Wyckoff sites. In half Heusler (HH) alloy (XYZ) 4d sites remain vacant, leading to 1:1:1 stoichiometry. Depending upon the site occupancy by elements, these HH alloys may exist in three different structural configurations (viz. the $\alpha$, $\beta$ and $\gamma$ phases). The details of the structures of the three phases are provided in the Fig.1 of section S1 of supplementary information. Thus, the identification of energetically the most favorable structural configuration comprises the first step. This is also one of the most lengthy steps, since for each of the LiY$_p$Y$^\prime_{1-p}$S compositions (Y, Y$^\prime$ = V, Cr, Mn, Fe, Co, and Ni; $p$ = 0, 0.25, 0.75, and
1), three different structural replicas (for the $\alpha$, $\beta$ and $\gamma$ phases) are considered while taking into account of spin polarization \ref{compu}.

\begin{figure}[t]
	\center
		\includegraphics[scale=0.34]{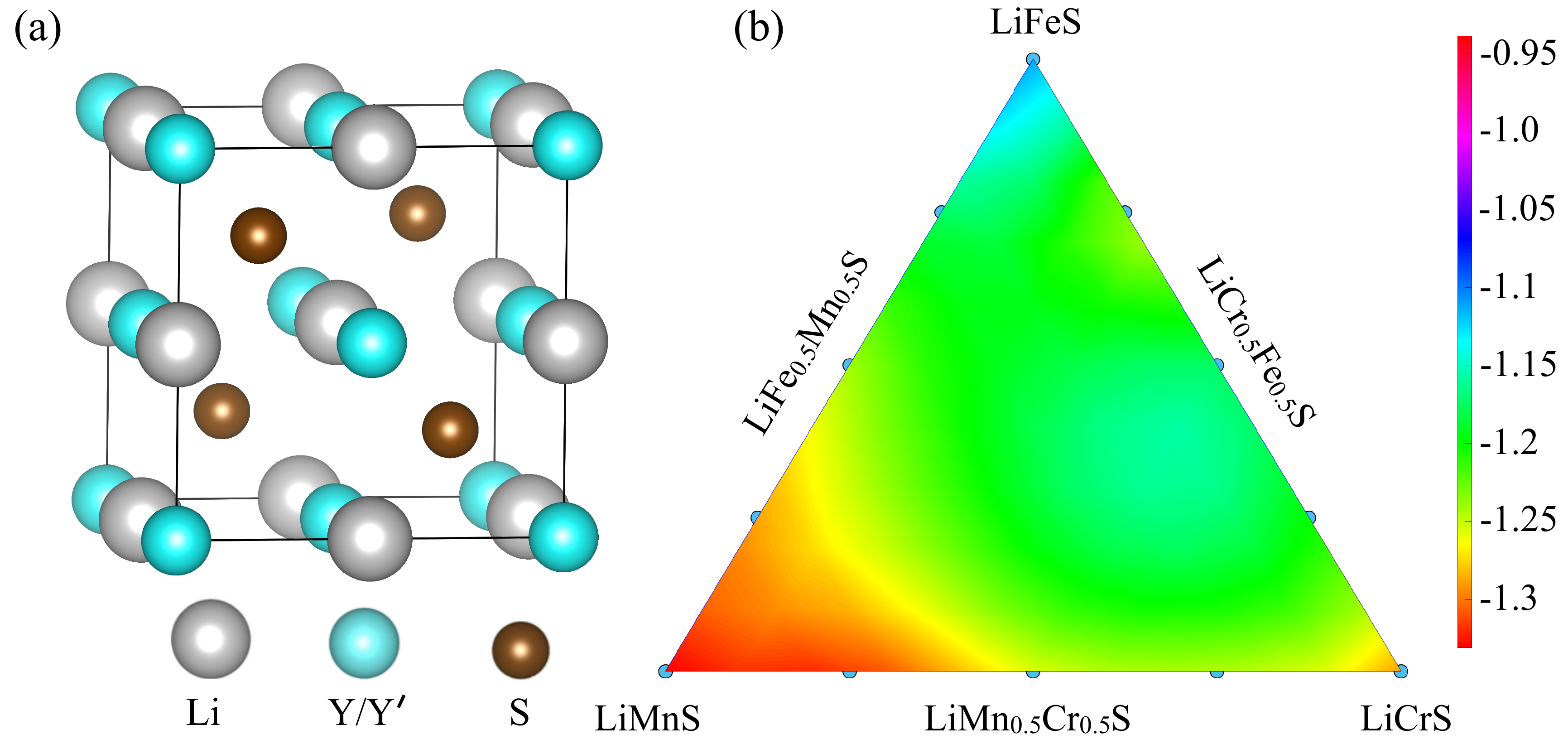}
		\caption{(a) The conventional cell of the most stable ($\beta$) phase of half Heusler composition. The Li and S atoms fully occupy the specific (4b and 4c) Wyckoff sites, while the 4a sites are partially occupied by the transition metal atoms (Y/Y$^\prime$). (b) The formation energy plot of few selected compositions originating from the parents LiMnS, LiFeS and LiCrS. The color scale shows the variation of the formation energy.}
		\label{uc}
\end{figure}

Once the ground state geometry is identified, we calculate their respective formation energy using;    

\begin{equation}
E_\mathrm{f} = E(\mathrm{LiY}_{p}\mathrm{Y^\prime}_{1-p}\mathrm{S}) - \mu_{\mathrm{Li}} - p \cdot \mu_{\mathrm{Y}} - (1-p) \cdot \mu_{\mathrm{Y^\prime}} - \mu_{\mathrm{S}} 
\end{equation}

whereby, E(LiY$_p$Y$^\prime_{1-p}$S) is the ground state total energy of the respective composition. $\mu_{\mathrm{Li}}$, $\mu_\mathrm{Y}$, $\mu_\mathrm{Y^\prime}$,
and $\mu_\mathrm{S}$ are the chemical potential of the Li, Y, Y$^\prime$ and S elements, which is taken from their stable earth abundant bulk phases (the details of the phases taken as reference is provided in the section S2 of the supplementary information). Thus, the compositions with a negative formation energy is retained for further analysis while the remaining ones are discarded. 

However, a negative formation energy does not always ensure the dynamical stability of the composition. On the otherhand, the dynamical stability of the composition is extremely important for ensuring the formation of the composition in laboratory. Thus, at this stage phonon calculations are performed under the framework of density functional perturbation theory for energetically the most stable compositions taking into account of spin polarization. The calculation settings are carefully chosen to ensure that the dynamical instabilities (if any) are not an artifact arising from the choice of cell (see section \ref{compu} for details). Thus, beyond this step only the dynamically stable compositions are retained for magnetic structure analysis. 

\begin{figure}[t]
	\center
	\includegraphics[scale=0.48]{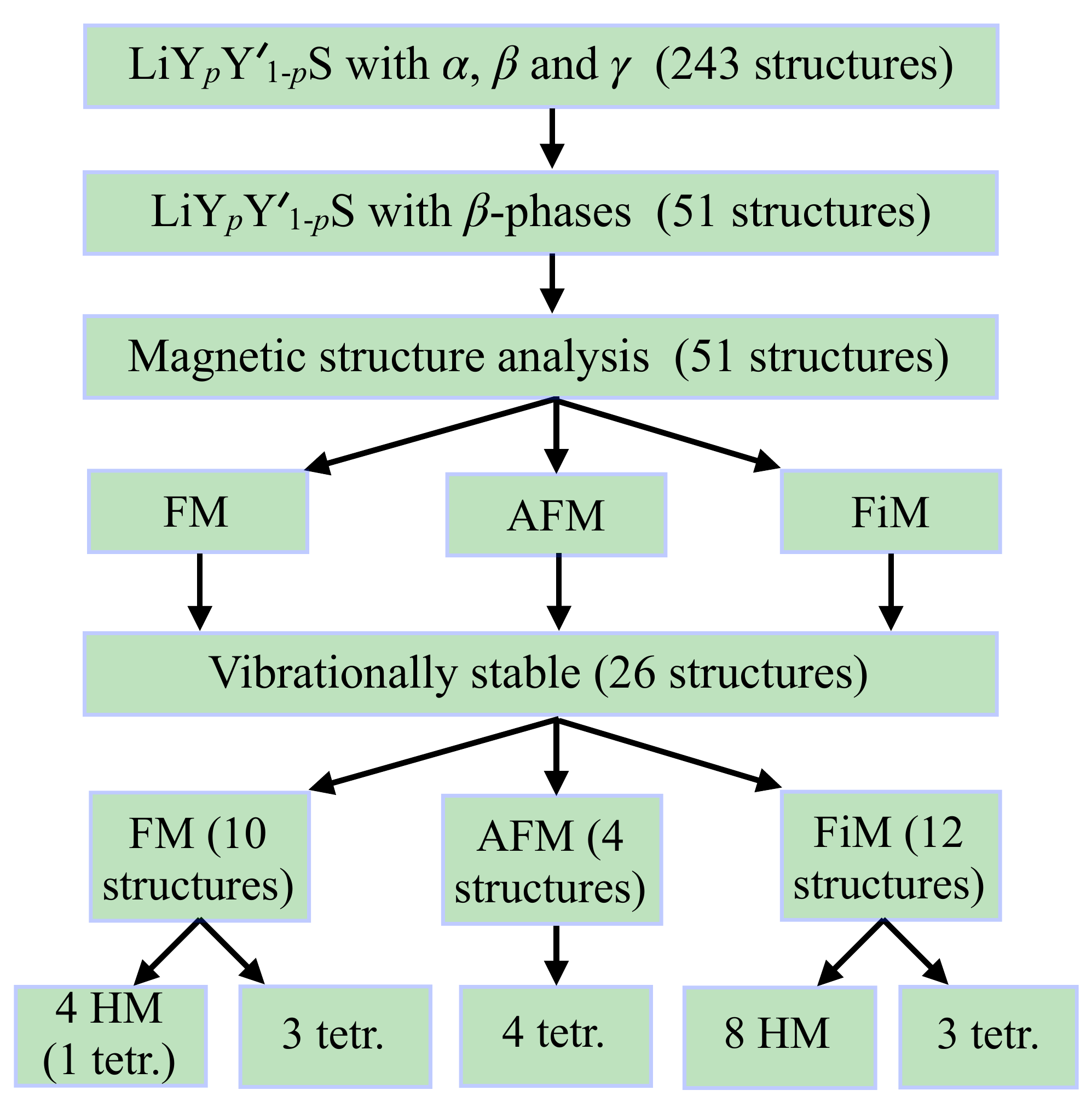}
	\caption{Flowchart of high throughput calculations showing the different steps of the screening criteria designed for selecting compositions based on the stability, vibrational, and magnetic analysis.}
	\label{flowchart}
\end{figure}

Further, the magnetic analysis is again explicity performed for the dynamically stable compounds by comparing their energies in ferromagnetic (FM), antiferromagentic (AFM) and ferrimagnetic (FiM) configurations. Thereby, the lowest energy configuration is identified as the magnetic ground state. Furthermore, the percentage of spin polarization (SP)\cite{soulen1998measuring} is calculated for the FM and FiM configurations, using the following relation;

\begin{equation}
\mathrm{SP}=\left(\frac{\mathrm{DOS}^{\uparrow}-\mathrm{DOS}^{\downarrow}}{\mathrm{DOS}^{\uparrow}+\mathrm{DOS}^{\downarrow}}\right)_{\mathrm{E_F}} \times 100 ~ \%
\end{equation}       

where, SP is the percentage of the ratio of difference in the density of majority spin ($\uparrow$) and minority spin ($\downarrow$) states and the total density of spin states at the Fermi level E$_\mathrm{F}$. The compositions with 100\% spin polarization (i.e., the Half metallic ones) are particularly interesting because of their applications in spintronics devices as spin filters, magnetic tunnel junction etc \cite{felser2013spintronics}.  

Finally, the magnetocrystalline anisotropy energy (MAE) is analyzed, which may predominantly arise from the structural anisotropy arising from the tetragonal distortion due to the elemental substitutions in the compositions\cite{felser2015basics, faleev2017origin}. The anisotropy constant K gives the estimation of the MAE in the crystal, which is calculated using; 
\begin{equation}
\mathrm{K}=\frac{\mathrm{E}^{\parallel}-\mathrm{E}^{\perp}}{\mathrm{V}}
\end{equation} 

where, $\mathrm{E}^{\parallel}$ and $\mathrm{E}^{\perp}$ are the total energies when the crystal is magnetized along the in-plane and out-of-plane directions respectively for a crystal of volume V. In order to calculate the $\mathrm{E}^{\parallel}$ and $\mathrm{E}^{\perp}$, the global spin quantization axis is varied along the three crystallographic orientation i.e. along the (100), (010), and (001) axis, while incorporating the effect of spin orbit coupling (SOC) and the total energy is compared for each case. Accordingly, the orientation with the lowest total energy is referred to as the easy axis for magnetic polarization. 

Thereby, a composition with high spin polarization as well as high MAE is concluded to be ideal for the spintronics applications\cite{faleev2017origin, faleev2017heusler}. All these steps involved in high throughput screening are schematically shown in Fig.~\ref{flowchart}.  

\section{Results}

\subsection{Structural and stability}

As discussed briefly in the method section \ref{metho}, a pristine half Heusler alloy may exists in three different configurations, i.e. the $\alpha$, $\beta$ and $\gamma$, depending  upon the site occupancy of the X, Y, and Z elements. The site preference may also change depending upon the substitution in LiY$_p$Y$^\prime_{1-p}$S (Y, Y$^\prime$ = V, Cr, Mn, Fe, Co, Ni and p = 0, 0.25, 0.5, 0.75, 1). Thus, for each of the different compositions all the three phases (i.e. the $\alpha$, $\beta$ and $\gamma$) are considered to generate the different structural replicas for the high throughput calculation. As for instance, for generating the LiY$_{0.25}$Y$^\prime_{0.75}$S (LiY$_{0.75}$Y$^\prime_{0.25}$S) structures in one of the given phase (say $\alpha$), the conventional lattice of LiY$^\prime$S (LiYS) is considered and one of the four Y$^\prime$ (Y) atom is substituted with one Y (Y$^\prime$) atom. However, in order to generate the intermediate configuration, LiY$_{0.5}$Y$^\prime_{0.5}$S, starting with LiY$_{0.25}$Y$^\prime_{0.75}$S, one of the three remaining Y$^\prime$ atoms is substituted with one Y atom iteratively and the lowest energy structure is identified. Same process is repeated for the $\beta$ and $\gamma$ configurations. Thus, for each pair of elements at the Y site, it took 15 different structural replicas to identify one stable configuration. Given that 15 different pairs of elements are considered for the Y site, 225 structural replicas were generated for LiY$_p$Y$^\prime_{1-p}$S and similarly 18 more for the parents (LiY$^\prime$S /LiYS). 

\begin{figure}[t]
	\center
	\includegraphics[scale=0.36]{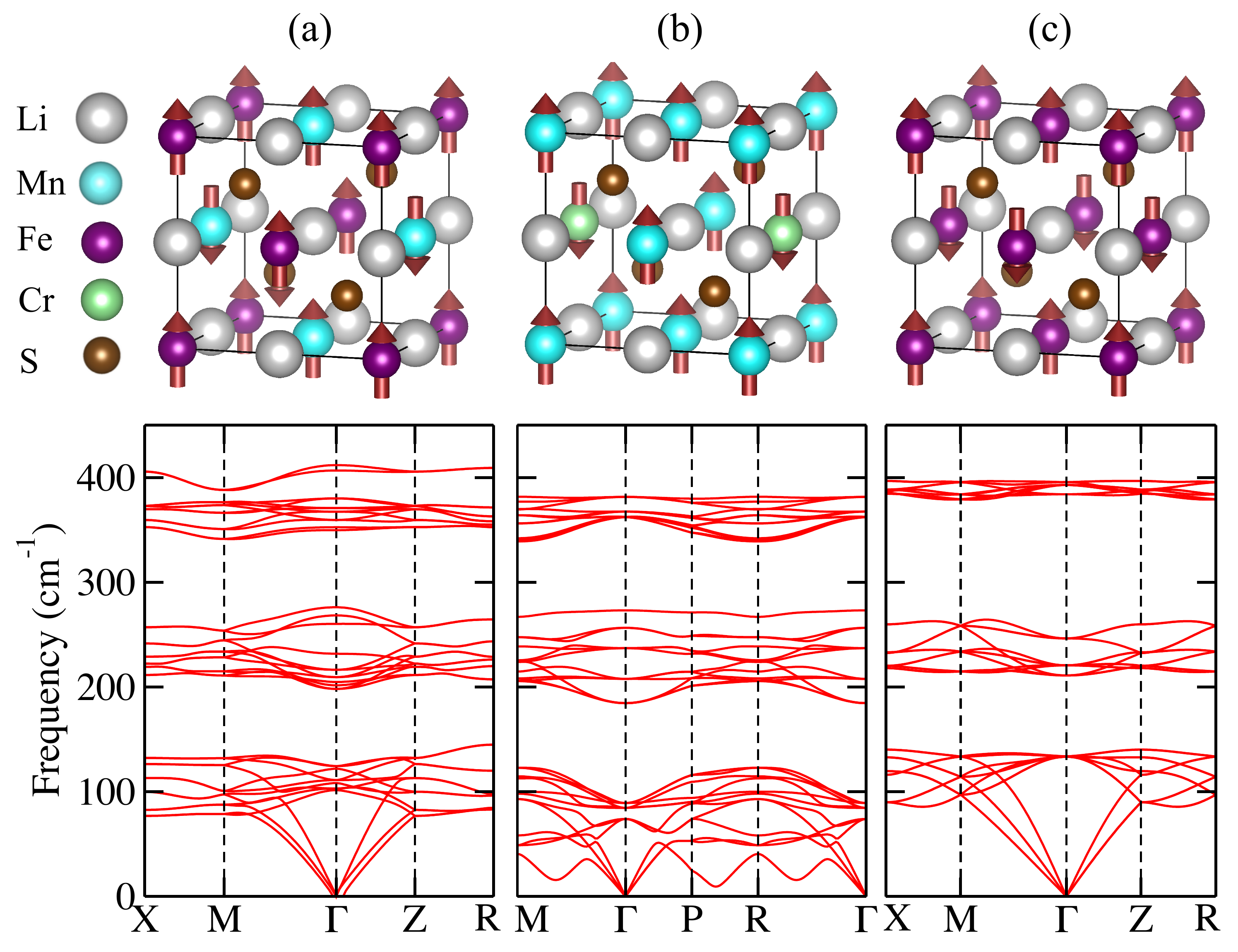}
	\caption{The ground state magnetic ordering (top panel) and the corresponding phonon band structure (bottom panel) of selective representative magnetic compositions viz.
		(a) LiFe$_{0.5}$Mn$_{0.5}$S (FM) (b) LiCr$_{0.25}$Mn$_{0.75}$S (FiM)  and (c) LiFeS (AFM). }
	\label{phonon}
\end{figure}

At the first step, spin polarized total energy calculations are performed on the structural replicas. The $\beta$ configuration, with X atom in the 4b, Y/Y$^\prime$ atom in the 4a and Z atom in the 4c position, is found to be more stable than the other two phases by more than 0.5 eV per formula unit (f.u.). Thereby, using the $\beta$ configurations for all the given compositions and the bulk energies of the constituent elements as reference, the formation energy of the compositions is calculated. The formation energy of few of the selected ones, generated from LiFeS, LiMnS and LiCrS as parents, is given in the three phase plot of Fig.~\ref{uc}(b). All compositions are found to have negative formation energy, while most of the compositions in the three phase diagram are found to have negative formation energy as compared to their parents i.e. LiMnS and LiCrS (shown in green), which indicates towards their formation in the laboratory.

It should be noted here that a spin polarized calculation may not be sufficient to predict the exact magnetic ground state of the composition. Since, the total energy contribution may marginally differ when different spin alignments (i.e. in the FM, FiM and AFM) are explicitly taken into consideration. Thus, for all of the 51 $\beta$ configurations, the magnetic ground state is calculated by varying the spin alignments in different FM, FiM and AFM spin configuration and thereby, lowest energy configuration is identified as true magnetic ground state of the composition (see Table.1 of section S3 of supplementary information for the formation energy and magnetic properties of all the stable compounds). Furthermore, we perform phonon calculations on these ground state magnetic structure and check their dynamical stability. Only 26 compositions are found to be dynamical stable (with no or very negligible negative phonon modes)(see Fig.~2 of section S4 of the supplementary information for the dynamical stability of all the FM compositions). These compositions are retained for further calculation of the \% of SP, MAE etc. 

\subsection{Magnetic properties}
The spin configuration of the dynamically stable ones reveals 10 FM, 12 FiM and 4 AFM compositions. We calculated the \% SP in FM and FiM ones at the \hl{Fermi level}, which is found to be 100\% in 4 FM and 8 FiM ones. The magnetic properties i.e. the type of magnetic configuration, \% SP and net magnetization per f.u. for all these dynamically stable compositions are provided in Table.1.

\begin{table}[htbp]
	\center
	\caption{Details of the structural and magnetic properties of the 26 dynamically stable compositions as obtained from high throughput search. The magnetic configuration (M$_{\mathrm{Conf.}}$), tetragonality (c/a), percentage of spin polarization (\% SP) and net magnetization (M) are enlisted for each of these compositions.}
	\label{tab:table1}
	\begin{tabular} {p{2.8cm}p{1.4cm}p{0.8cm}p{1.2cm}p{1.3cm}}  
		\hline
		\hline
		\textbf{Compositions} & \textbf{ M$_{\mathrm{Conf.}}$} & \textbf{M} & \textbf{SP} & \textbf{c/a} \\
		\hline
		
		LiFeS  & AFM & 0 & 0 & 0.994 \\
		
		LiCoS  & FM  & 1.99 & 89 & 1\\
		
		LiNiS  & FM & 0.39 & 59 & 1\\
		
		LiCo$_{0.5}$Mn$_{0.5}$S  & FM & 2.99 & 96 & 1.009\\
		
		LiFe$_{0.5}$Co$_{0.5}$S  & FM & 2.36 & 54 & 0.998\\
		
		LiFe$_{0.5}$Mn$_{0.5}$S  & FM & 3.50 & 100 & 1.011\\
		
		LiCr$_{0.5}$Fe$_{0.5}$S  & FiM & 0.99 & 84 & 1.014\\
		
		LiCr$_{0.5}$Mn$_{0.5}$S  & FiM & 0.49 & 72 & 1.004\\
		
		LiFe$_{0.5}$V$_{0.5}$S  & FiM & 0.38 & 60 & 1.01\\
		
		LiCo$_{0.5}$Ni$_{0.5}$S  & AFM & 0 & 0 & 1.002\\
		
		LiCo$_{0.25}$Ni$_{0.75}$S  & FM & 0.57 & 37  & 1\\
		
		LiFe$_{0.25}$Co$_{0.75}$S  & FM & 2.22 & 100  & 1\\
		
		LiFe$_{0.25}$Ni$_{0.75}$S  & FM & 0.90 & 43  & 1\\
		
		LiCr$_{0.25}$Fe$_{0.75}$S  & FiM & 1 & 100  & 1\\
		
		LiCr$_{0.25}$Mn$_{0.75}$S  & FiM & 1.75 & 100  & 1\\
		
		LiV$_{0.25}$Co$_{0.75}$S  & FiM & 0.5 & 100  & 1\\
		
		LiV$_{0.25}$Fe$_{0.75}$S  & FiM & 1.25 & 100  & 1\\
		
		LiV$_{0.25}$Mn$_{0.75}$S  & FiM & 2 & 100  & 1\\
		
		LiV$_{0.25}$Ni$_{0.75}$S  & FiM & 0.25 & 100  & 1\\
		
		LiCo$_{0.25}$Mn$_{0.75}$S  & AFM & 0 & 0 & 1.009\\
		
		LiFe$_{0.75}$Co$_{0.25}$S  & FM & 2.71 & 100  & 1\\
		
		LiFe$_{0.75}$Mn$_{0.25}$S  & FM & 3.25 & 100  & 1\\
		
		LiCo$_{0.75}$Mn$_{0.25}$S  & FiM & 0.08 & 100  & 1\\
		
		LiCo$_{0.75}$Ni$_{0.25}$S  & FiM & 1.65 & 47  & 1\\
		
		LiNi$_{0.75}$Mn$_{0.25}$S  & FiM & 0.75 & 100  & 1\\
		
		LiFe$_{0.75}$Ni$_{0.25}$S  & AFM & 0 & 0 & 1.001\\
		\hline
		\hline
		
	\end{tabular}
\end{table}

\begin{figure}[t]
	\center
	\includegraphics[scale=0.34]{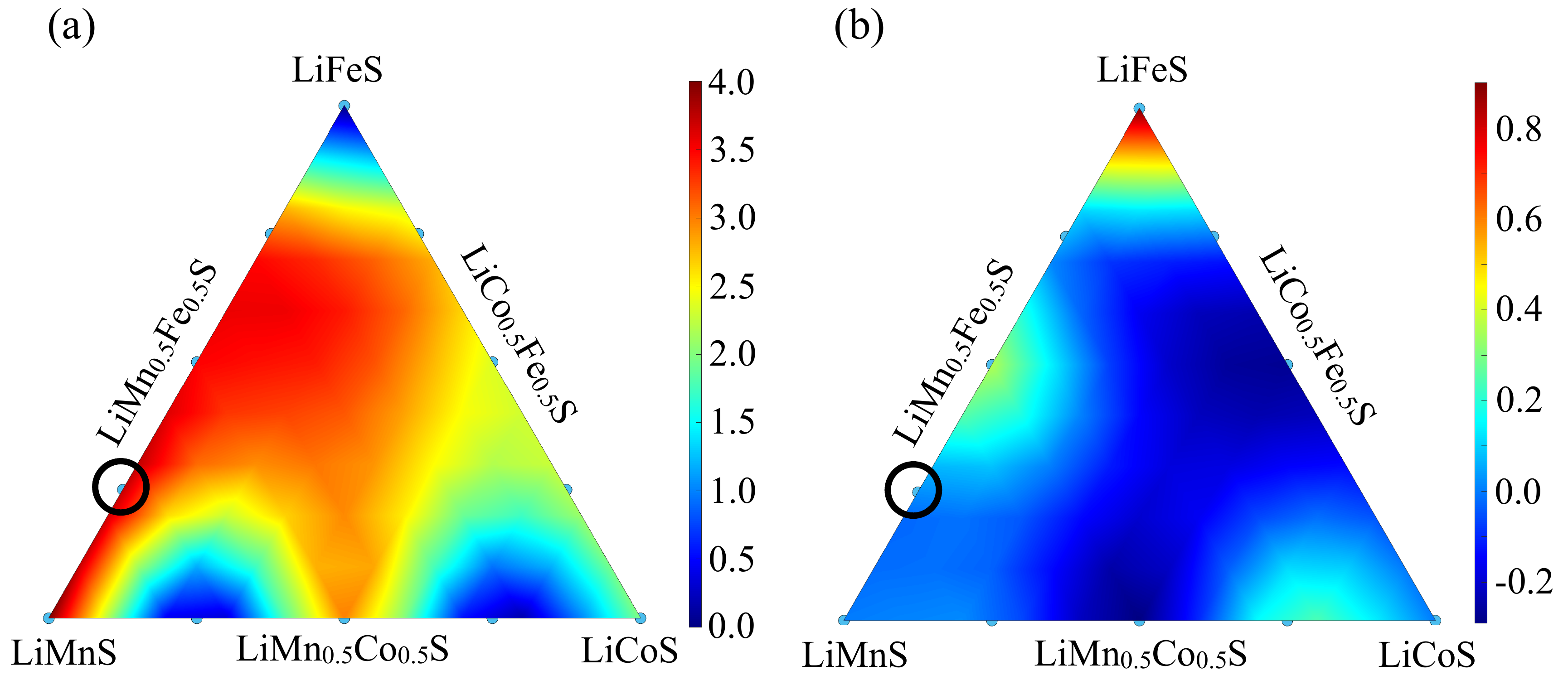}
	\caption{The total (a) magnetization ($\mathrm{\mu_B}$/f.u.) and (b) magnetocrystalline anisotropy constant K (MJ/$\mathrm{m^3}$) plot of few selected compositions originating from the parents LiMnS, LiFeS and LiCoS. The color scale shows the variation in respective magnitudes.}
	\label{mag-mae}
\end{figure}

Finally, we also check for tetragonal distortion in the structure of the magnetic compositions. 10 of these structures are found to be tetragonal. These are further taken in consideration for explicit calculation of the magnetocrystalline anisotropy energy (MAE). These \hl{intermediate} compositions (see Table \ref{tab:table1}) are found to show a wide variety of magnetic interactions including several exotic properties such as half metallicity and MAE. The net magnetization is found to be dominant in Mn-Fe based ones (with highest of 3.5 $\mu_B$ in LiFe$_{0.5}$Mn$_{0.5}$S) followed by Mn-Co (with highest of 2.99 $\mu_B$ in LiMn$_{0.5}$Co$_{0.5}$S) and Fe-Co ones (with highest of 2.71 $\mu_B$ in LiFe$_{0.75}$Co$_{0.25}$S). Several compositions are also found to show half metallicity (i.e. almost 100\% spin polarization) while possessing high moments per f.u. of $>$ 2 $mu_B$ viz. LiFe$_{0.5}$Mn$_{0.5}$S, LiCo$_{0.5}$Mn$_{0.5}$S, LiFe$_{0.25}$Co$_{0.75}$S, LiV$_{0.25}$Mn$_{0.75}$S, LiFe$_{0.75}$Co$_{0.25}$S and LiFe$_{0.75}$Mn$_{0.25}$S. Finally, some of these compositions are also found to be tetragonal viz. LiFe$_{0.5}$Mn$_{0.5}$S and LiCo$_{0.5}$Mn$_{0.5}$S. The net magnetization and the corresponding MAE is provided in Fig.~\ref{mag-mae} for a few selected compositions with interesting magnetic properties viz. the ones generated from LiMnS, LiFeS and LiCoS as parents. The compositions having both Mn and Fe are found to be highly magnetic (as shown in Fig.~\ref{mag-mae} a), while the Fe and Co/~Mn are found to show high in-plane magnetocrystalline anisotropy (as shown in the dark blue region in Fig.~\ref{mag-mae} b). 

\begin{figure}[t]
	\center
	\includegraphics[scale=0.476]{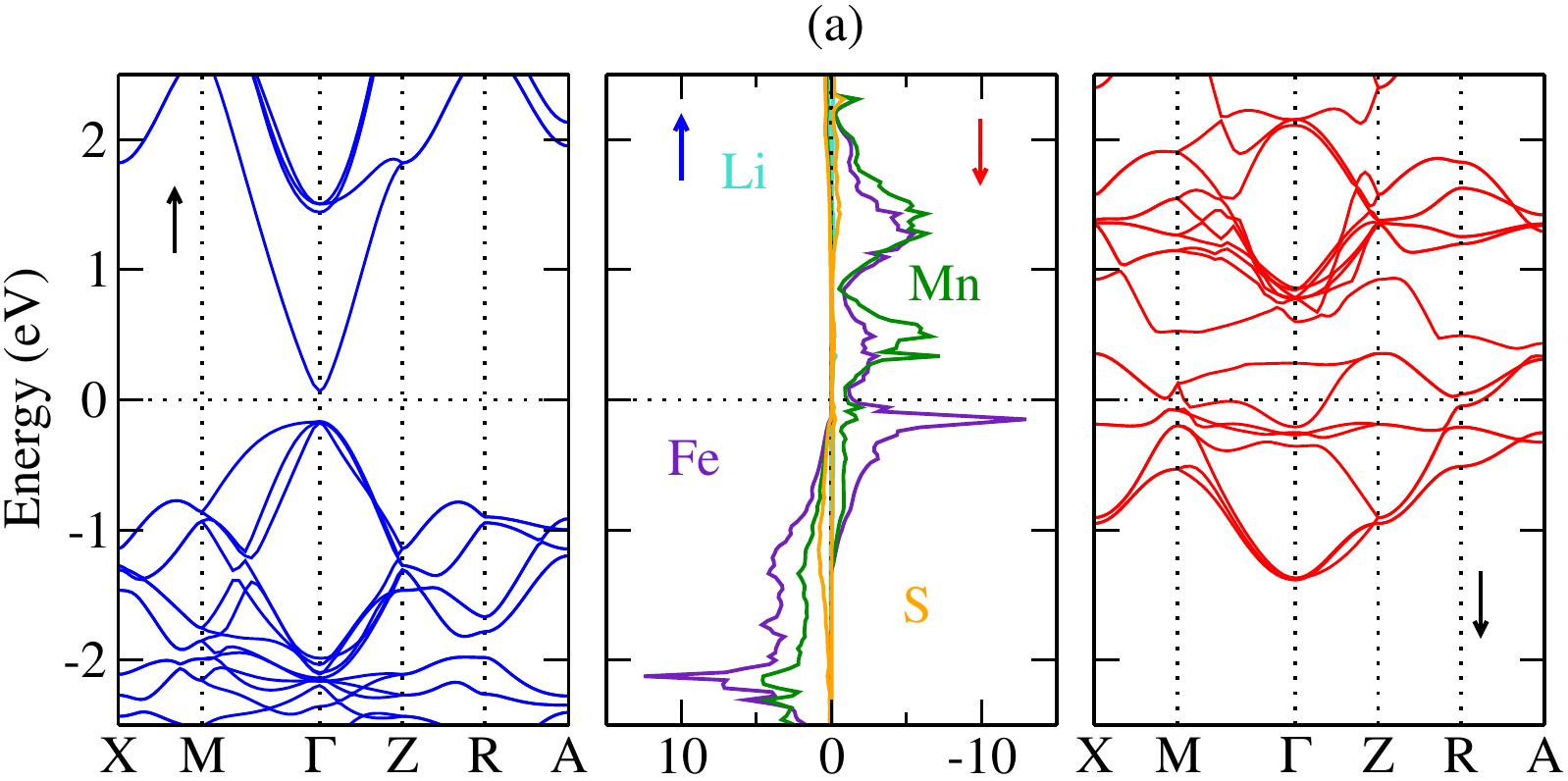}
	\includegraphics[scale=0.476]{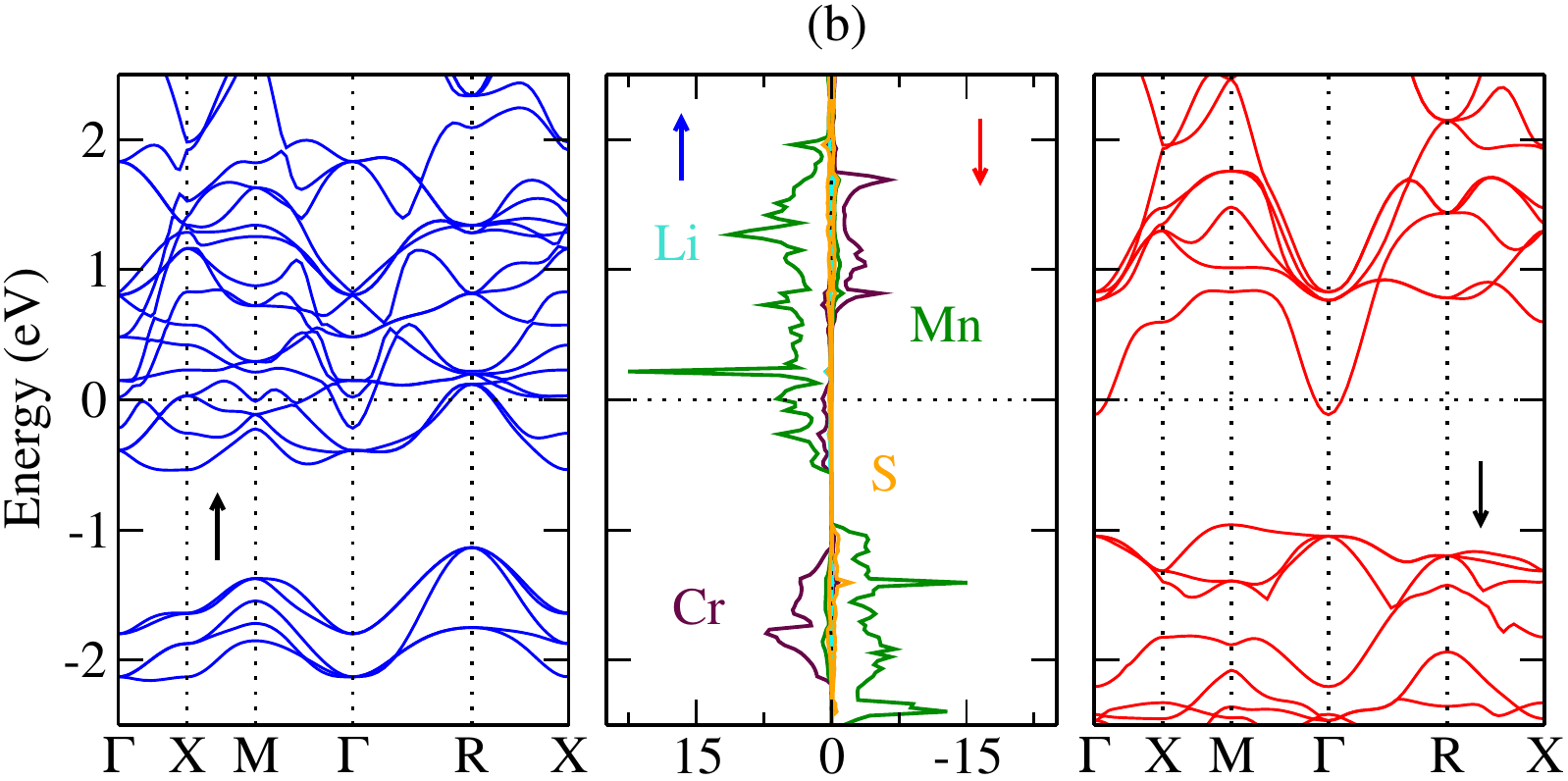}
	\includegraphics[scale=0.476]{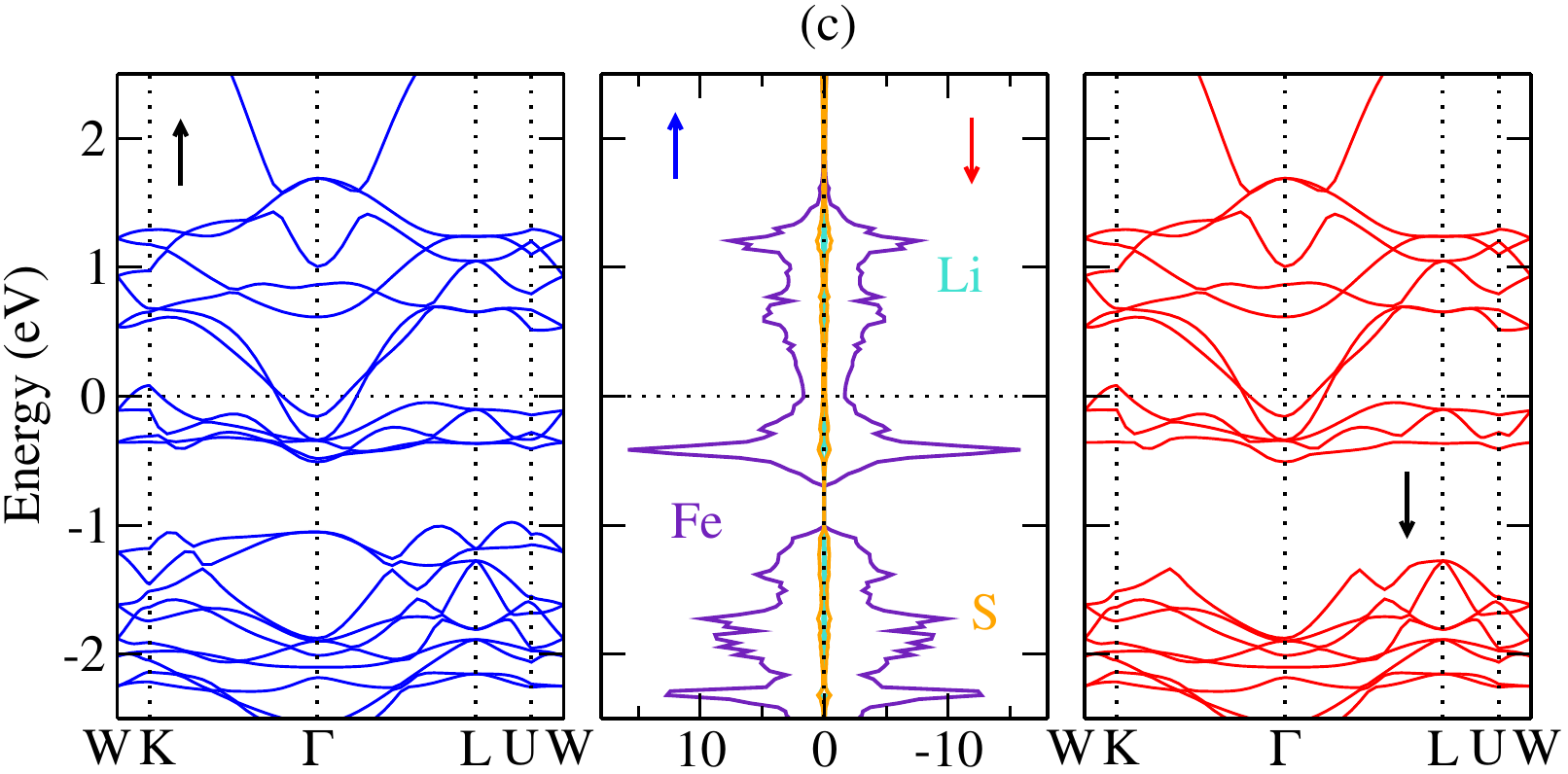}
	\caption{Electronic band structure of up (left) and down (right) spin components and their corresponding density of states (middle) of one of the representative composition with ferro, ferri and anti ferro ordering viz. (a) LiFe$_{0.5}$Mn$_{0.5}$S, (b) LiCr$_{0.25}$Mn$_{0.75}$S, and (c) LiFeS.}
	\label{dosband}
\end{figure}

The electronic band structure and density of states plots for three representative compositions i.e. one from the FM, FiM and AFM category respectively is shown in Fig.~\ref{dosband}. LiFe$_{0.5}$Mn$_{0.5}$S (Fig.~\ref{dosband} a) is found to be half metallic FM with 100\% spin polarization, since only the down spin channel is found to be present at the \hl{Fermi level}. Here, the spin on both the Mn (with moment of 3.8 $\mu_B$ per atom) as well as the Fe (with moment of 2.7 $\mu_B$ per atom) atoms are found to be alligned in the same direction, which gives rise to a net uncompensated moment of 3.5 $\mu_B$ per f.u. Intrestingly, the magnetic moment on Fe atoms in LiFe$_{0.5}$Mn$_{0.5}$S is found to be higher as compared to that in the pure bulk Fe (2.2 $\mu_B$). The distance between the magnetic atoms in this composition is compared with its parents. Fe-Fe neighbor distance (3.864 \AA) is found to be the lowest in the AFM coupled parent LiFeS alloy, whereas the Mn-Mn neighbor distance (4.007 \AA) is found to be maximum in the FM coupled parent alloy LiMnS. The Fe-Mn (3.953 \AA) neighbor distance in LiFe$_{0.5}$Mn$_{0.5}$S is found to be intermediate to those in the two parents. This is also in accordance with the Bethe Slater curve for transition metals \cite{cardias2017bethe}, which states that the magnetic interaction in a given alloy varies as a function of neighbor distance and d orbital contribution. A bandgap of $\approx$ 0.23 eV is observed in up spin channel of LiFe$_{0.5}$Mn$_{0.5}$S. To understand the origin of the half metallicity, the orbital projected band structure of LiFe$_{0.5}$Mn$_{0.5}$S composition is analysed using PyProcar\cite{herath2020pyprocar}. The projection of the d orbitals (t$_{2g}$ and e$_{g}$ individually) on the electronic band structure is analyzed for the spin up and spin down channels separately. For the spin up channel, the valence band maximum is found to be formed from the hybridization of t$_{2g}$ orbitals of Mn and Fe (three orbitals converging at the valence band maximum highlighted by red color in the left column of Fig. 6). In the conduction band minimum of the spin up channel, no significant contribution is observed from the d orbitals. For the spin-down channel, the t$_{2g}$ and e$_{g}$ orbitals of Mn and t$_{2g}$ of Fe are not found to make any significant contribution close to the \hl{Fermi level}, while the e$_{g}$ orbitals of Fe (shown in green just below the Fermi level in the top right column of Fig. 6) makes significant contribution near the Fermi level. The down spin is found to be completely metallic giving rise to the half metallic ferromagnetic characteristic of the composition. \hl{This composition also showed a fairly high MAE of 0.1 meV per formula unit which corrsponds to the anisotropy constant of 0.36 MJ/m$^3$} (see Fig.\ref{mag-mae} b) and the (001) axis is found to be the easy axis of magnetic polarization for FM spin alignment. However, the orbital contribution to magnetism is found to be significantly low in both Mn (0.035 $\mu_B$) as well Fe (0.11 $\mu_B$) atom as compared to the dominant spin part, which comprised the majority of the moment on each of the atom. 

\begin{figure}[t]
	\center
	\includegraphics[scale=0.25]{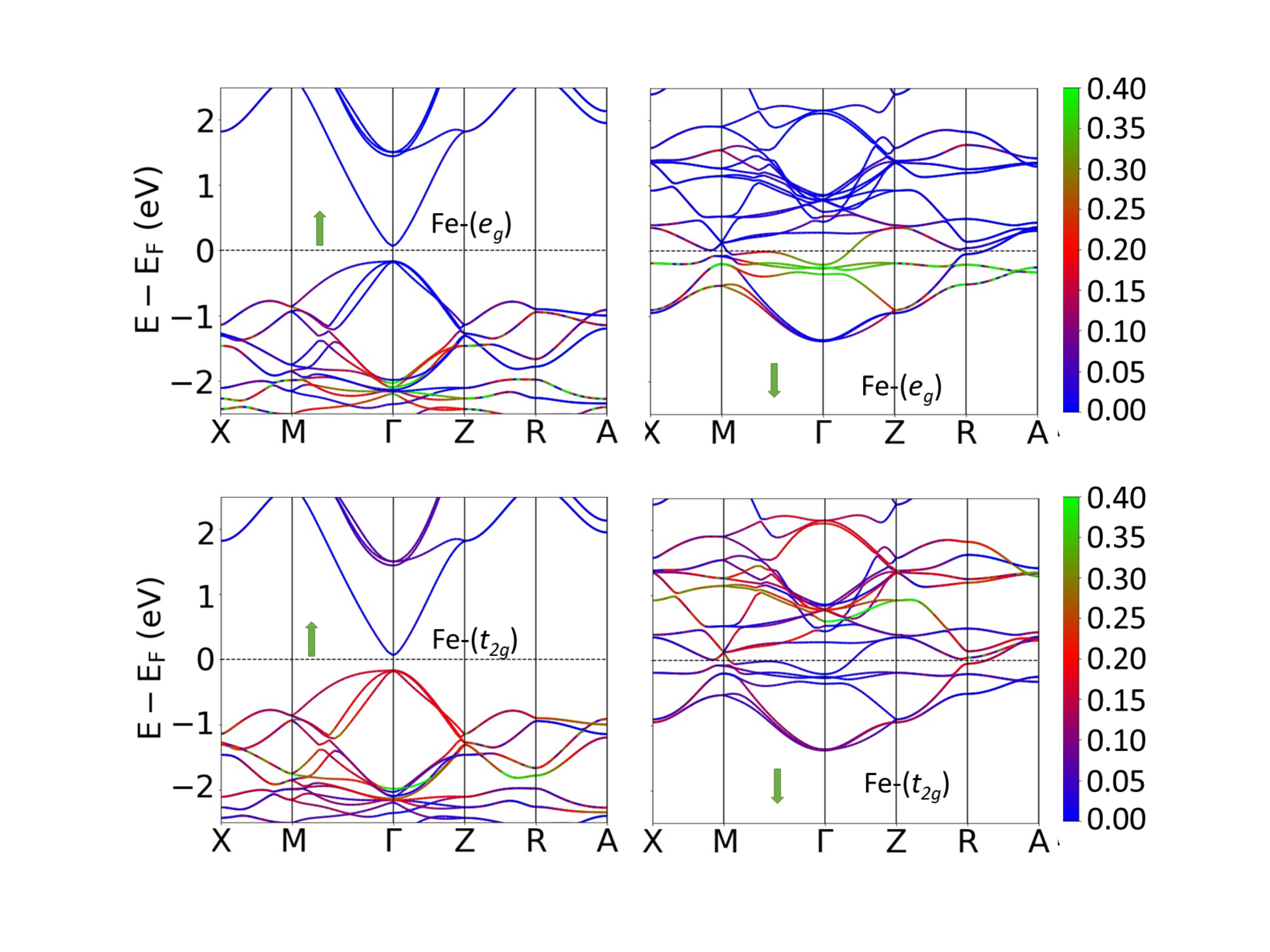}
	\includegraphics[scale=0.25]{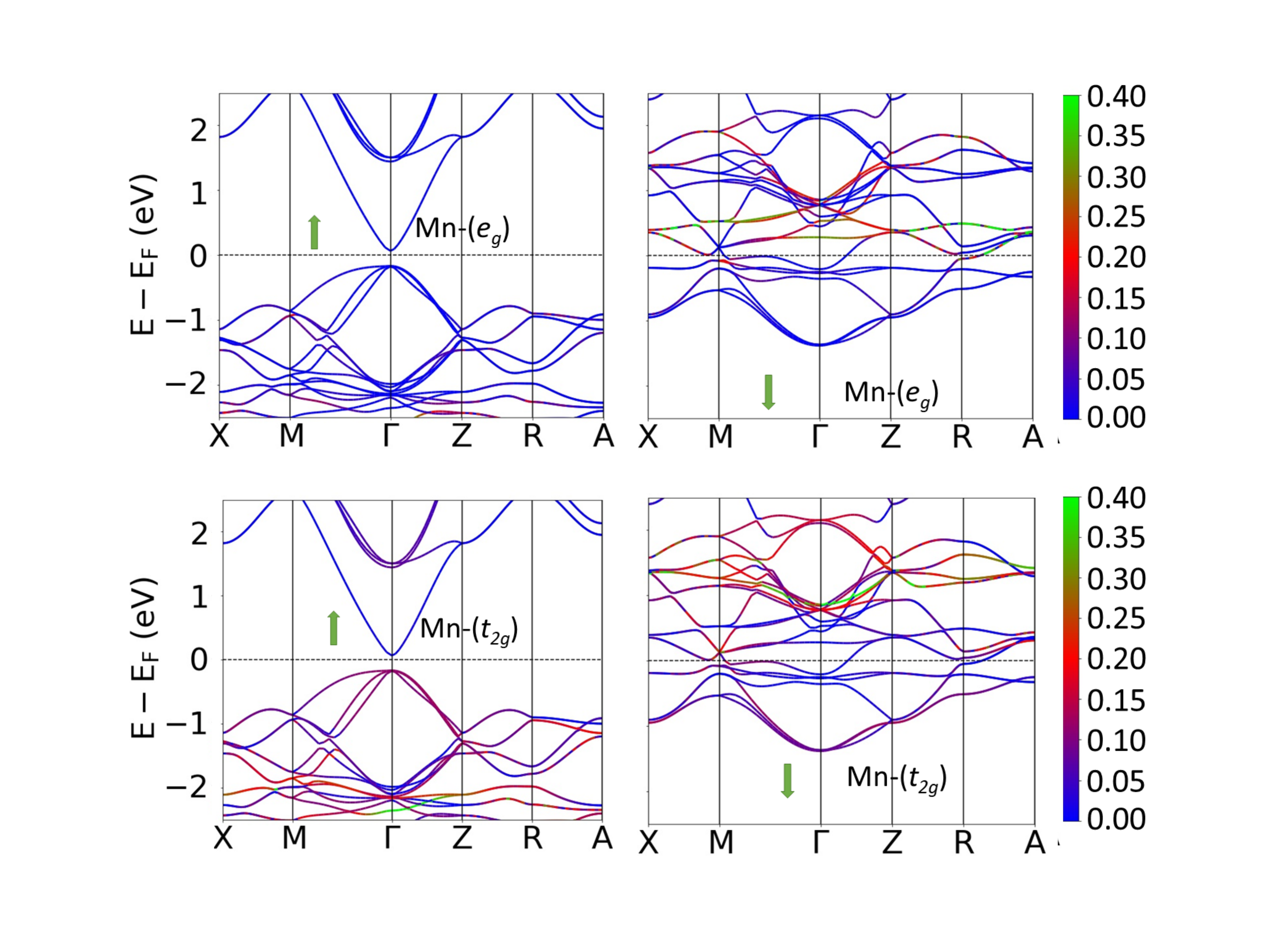}
	\caption{The t$_{2g}$ and e$_{g}$ orbital projections of the magnetic elements (Mn and Fe) in the ferromagnetic composition LiFe$_{0.5}$Mn$_{0.5}$S
		for the up spin (left) and down spin (right). The color scale denotes the weighted contribution stemming from the particular orbital.}
	\label{partial}
\end{figure}

Our high throughput search led to the identification of several FiM compositions. Out of these, few were also found to be half metallic with high anisotropy. One of such ferrimagnetic (FiM) compositions viz. LiCr$_{0.25}$Mn$_{0.75}$S is analyzed and presented as a representative case. It is found to be a FiM half metal with significant up spin contribution and almost negligible down spin contribution at the Fermi level (see Fig.~\ref{dosband} (b)). The moment on Cr atom (3.71 $\mu_B$ per atom) is found to be higher, but opposite to the moment on Mn atom (3.5 $\mu_B$ per atom), which led to a net moment of 1.75 $\mu_B$ per f.u (as shown in Fig.~\ref{phonon} b). Further, it is also found to be anisotropic with (001) is an easy axis of FiM polarization with \hl{MAE of 0.017 meV per formula unit and the corresponding anisotropy constant of 0.134 MJ/m$^3$ }(as shown in Fig. \ref{mag-mae} b). As expected the orbital contribution to the moment of the Cr and Mn atoms is found to be fairly lower (of 0.012 $\mu_B$ and 0.055 $\mu_B$ respectively) as compared to the spin moment (of 3.71 $\mu_B$ and 3.5 $\mu_B$ respectively). Despite of its FiM magnetic configuration, it is found to show a good magnetization of 1.75 $\mu_B$, with an out of plane MAE, making it a potential candidate for the spintronics application.

AFM alloys with high MAE also have important spintronics applications viz. in spin transfer torque (STT) based magnetic random access memory (MRAM) \cite{zhuravlev2018perpendicular, wadley2016electrical} etc \cite{welbourne2021antiferromagnetic}. Thus, as a representative case, the LiFeS alloy is analyzed and presented. In its stable AFM configuration, Fe atoms in every plane is found to possess equal (of 2.4 $\mu_B$) and unidirectional moments, which were cancelled by the moments of Fe atoms in adjacent plane (as shown in Fig. \ref{phonon} c). This is confirmed by the finite, but equal and opposite contribution of the spin up and spin down channels at the \hl{Fermi level} (as shown in Fig. \ref{dosband} c). The (001) axis is found to be the easy axis with \hl{MAE of 0.231 meV per formula unit and an anisotropy constant of 0.902 MJ/m$^3$.} The spin and orbital moment on the Fe atoms are found to be 2.40 $\mu_B$ and 0.096 $\mu_B$ respectively.  

The Curie temperature (T$_\mathrm{c}$) is also another important property for magnetic application of an alloy and hence, is discussed hereby qualitatively. T$_\mathrm{c}$ of an alloy mainly depends upon the interaction strength between the magnetic ions. Thereby, in another words, T$_\mathrm{c}$ can be qualitatively predicted from the elemental constituents and their uncompensated atomic magnetic moment in the solid. Among the elemental solids, Fe, Co, and Ni are already known to show high T$_\mathrm{c}$. Most of magnetic compositions, enlisted in Table \ref{tab:table1}, also possess one of these elements as constituent. Also, it is generally seen that in half Heusler alloys, with high magnetic moment ($>$ 3 $mu_B$), T$_c$ tends to be above room temperature \cite{zhang2016half, wurmehl2006half}. This is in agreement with the T$_c$ calculated using Monte Carlo Simulation by Sattar \textit{et al.} on alkali metal based half Heusler alloys of formula XYZ, where X is alkali metal, Y is a 3d TM element and Z is a chalcogen element. They found that T$_c$ in these compositions is generally higher than 300 K Thus, from all these literature a crude idea on T$_c$ of LiY$_p$Y$^\prime_{1-p}$S (Y, Y$^\prime$ = V, Cr, Mn, Fe, Co, Ni and p = 0, 0.25, 0.5, 0.75, 1) compositions can be drawn to be above room temperature. 

\section{Conclusions}
We employ first principles DFT calculations to design high throughput loops to screen stable half Heusler (HH) compositions viz. LiY$_p$Y$^\prime_{1-p}$S (Y, Y$^\prime$ = V, Cr, Mn, Fe, Co, Ni and $p$ = 0, 0.25, 0.5, 0.75, 1) for spintronics applications. Taking into account of all possible geometric configurations arising from the site preference of the elements, a pool of 243 different structures is generated. By comparing their total energy obtained after incorporating the contribution stemming from the spin polarization, 51 low energy structures (belonging to the $\beta$ phase) are selected. As this stage, the ground state energy of each of these compositions is compared for different spin alignments explicitly (ie. for ferro, ferri and anti ferro) and the lowest energy structure is referred to be the magnetic ground state. An accurate phonon calculation is performed and 26 of the dynamically stable ones are retained. 10 FM, 12 FiM and 4 AFM compositions are screened to be dynamically stable, out of which 4 half metallic FM and 8 half metallic FiM are identified. A tetragonal distortion is observed in 3 FM (1 half metal), 3 FiM and 4 AFM leading to a magneto crystalline anisotropy (MAE) in the system. The ferromagnetic  LiFe$_{0.5}$Mn$_{0.5}$S composition is found to have the highest easy-axis anisotropy constant of 0.36 MJ/$m^3$ with 100\% spin-polarization with net magnetization of 3.5 $\mu_B$. The rest of the magnetic alloys are found to possess an anisotropy constant in the range of 0.001 MJ/$m^3$ $<$ K $<$ 0.36 MJ/$m^3$. The present work provides a blueprint for the range of magnetization, magneto crystalline anisotropy and spin polarization that can be achieved in LiY$_p$Y$^\prime_{1-p}$S compositions. 

\section{Computational details}
\label{compu}
We performed first-principles density functional theory calculation within projector augmented wave (PAW) method based on the plane-wave basis set as employed in Vienna ab initio simulation package (VASP)\cite{hafner1997vienna, hafner2008ab, blochl1994projector, kresse1999ultrasoft, kresse1996efficient, kresse1996efficiency}. The PBE (Perdew, Burke, and Ernzerhof) corrected GGA (generalized gradient approximation) energy functional was used addressing the exchange and correlation interactions\cite{perdew1996generalized}. The PAW potentials \cite{kresse1999ultrasoft, blochl1994projector}for Li, Y, Y$^\prime$, and S have been used to manifest the ion core and valence electron interaction. The kinetic energy cutoff for plane waves was set to 500 eV for the calculations. The optimized lattice constants were obtained by performing polarized spin calculations and relaxing the conventional unit cell within the conjugate gradient method. The structure was relaxed till forces on each ion were less than 0.001 eV/\AA, and total energy is converged within the convergence limit of $10^{-5}$ eV. For each of the different structures, the k-mesh was employed by checking in the total energy convergence. \hl{For calculating the magnetic anisotropy, the same k-mesh size was employed which was used for achieving the convergence in the total energy of the structures. In contrast, for DOS calculations, a denser k-mesh of $27\times 27\times 27$ is employed.} For non-cubic structures, the k-points were chosen as per the inverse proportional ratio of the lattice constants. The spin-orbit coupling was included in the total Hamiltonian while doing magnetocrystalline anisotropy calculation as it is fundamental origin of the anisotropy. Phonon band structure was plotted using the phonopy code based on the density functional perturbation theory \cite{togo2010first, gonze1997dynamical, chaput2011phonon}. The thermodynamic properties were obtained by using the phonon frequencies from the harmonic band structure.

\section*{Acknowledgements}

AB acknowledges the funding received as IIT B seed grant project (RD/0517-IRCCSH0-043), SERB ECRA project (ECR/2018/002356) and BRNS regular grant (BRNS/37098). The high performance computational facilities viz. Aron (AbCMS lab, IITB), Dendrite (MEMS dept., IITB), and Spacetime, IITB and CDAC Pune (Param Yuva-II) are acknowledged for providing the computational hours.

\section*{References}

\bibliography{mybibfile}

\end{document}